*Review Article*

# Factors Affecting Terahertz Emission from InGaN Quantum Wells under Ultrafast Excitation

**Muhammad Farooq Saleem** [ID],[1] **Ghulam Abbas Ashraf** [ID],[2] **Muhammad Faisal Iqbal** [ID],[3] **Rashid Khan** [ID],[4] **Muhammad Javid** [ID],[5] **and Tianwu Wang** [ID][1]

[1]*GBA Branch of Aerospace Information Research Institute, Chinese Academy of Sciences, Guangzhou 510530, China*
[2]*Department of Physics, Zhejiang Normal University, Jinhua 321004, Zhejiang, China*
[3]*Department of Physics, Riphah International University, Faisalabad 38000, Pakistan*
[4]*Zhejiang Provincial Key Laboratory of Advanced Chemical Engineering Manufacture Technology,
College of Chemical and Biological Engineering, Zhejiang University, Hangzhou 310027, China*
[5]*Institute of Advanced Magnetic Materials, College of Materials and Environmental Engineering, Hangzhou Dianzi University,
Hangzhou 310012, China*

Correspondence should be addressed to Muhammad Javid; javidssp908@gmial.com and Tianwu Wang; wangtw@aircas.ac.cn

InGaN quantum wells (QWs) grown on c-plane sapphire substrate experience strain due to the lattice mismatch. The strain generates a strong piezoelectric field in QWs that contributes to THz emission under ultrafast excitation. Physical parameters such as QW width, period number, and Indium concentration can affect the strength of the piezoelectric field and result in THz emission. Experimental parameters such as pump fluence, laser energy, excitation power, pump polarization angle, and incident angle can be tuned to further optimize the THz emission. This review summarizes the effects of physical and experimental parameters of THz emission on InGaN QWs. Comparison and relationship between photoluminescence properties and THz emission in QWs are given, which further explains the origin of THz emission in InGaN QWs.

## 1. Introduction

With the advent of THz radiation-based spectroscopic techniques, efficient THz generation has become an urgent need [1–4]. The THz spectroscopy has applications in environmental monitoring [5], imaging [6], biomedical diagnosis [7], material characterization [8], food inspection [9], medicine inspection [10], communication technologies [11], detection of explosives [12, 13], etc. The THz generation has been observed upon ultrafast excitation of semiconductors, alloys, gas plasmas, and some material combinations including LiNbO₃ [14], ZnTe [15–17], GaAs [18], InAs [2, 19], W/Co₄₀Fe₄₀B₂₀/Pt [20–23], InGaN [24], and quantum wells (QWs) [25–33], to name a few. The THz pulse generation is accomplished using various methods including photocurrent transients in gas plasmas, photocurrent surge from electro-optic materials [34–38], Cerenkov radiation in ferroelectric materials, difference frequency generation in nonlinear materials, spintronic emission from magnetic metal multilayers [23], and dynamic screening of electrostatic field in QWs [25, 27, 29, 30, 32, 33, 39–41]. Due to the extensive research on QWs for their importance in lighting applications, the growth technologies of QWs are mature and the rich underlying physics of QWs have been widely explored over the last few decades. One of the unique characteristics of InGaN/GaN QWs is the built-in electric field (of the order of MV/cm) originating from spontaneous and piezoelectric polarization in QWs grown along the [0001] crystal orientation of (c-plane) sapphire [42, 43]. The lattice mismatch between the QW and barrier materials results in a strain that causes piezoelectric polarization in QWs [44, 45]. The strong built-in piezoelectric fields result in a quantum confined stark effect (QCSE) [46, 47]. QCSE is the band-structure tilt in QWs due to the strong built-in piezoelectric field [48–51]. It leads to low optical transition energy and poor recombination efficiency in QW-LEDs as the wave function overlap decreases under the influence of this field [25, 52–57]. This built-in field however plays a central role in terahertz generation upon ultrafast excitation of InGaN/ GaN QWs [33]. The first photon from an ultrafast laser source increases the overlap of the electron-hole wave function and absorption coefficient for the subsequent photon [58]. The instantaneous absorption of the photon results in the spatial separation and the creation of polarized electron-hole pairs in a time interval much shorter than the typical recombination time [31]. The electric field of such a dipole has a polarity opposite to the built-in piezoelectric field. Due to the opposite polarity of these dipoles, the partial screening of the built-in field occurs, which leads to a local microscopic modification of both the optical and electronic properties of the system. The sufficient density of these dipoles excited by the ultrafast laser can cause complete screening of the built-in field. This process is known as "dynamical screening" in QWs. The screening effect induced by the carriers excited in the spatially separated states modifies the band structure, absorption coefficient, and electron-hole wave function within the duration of pulse excitation. The screening of the piezoelectric field results in THz emission. The effect was explained by considering a nanoscale capacitor that releases the stored electrostatic energy. The energy stored in QWs is released via THz emission under the influence of ultrafast excitation. These nonlinear effects can be observed since the time scale is much shorter than the typical recombination time of the carriers [58].

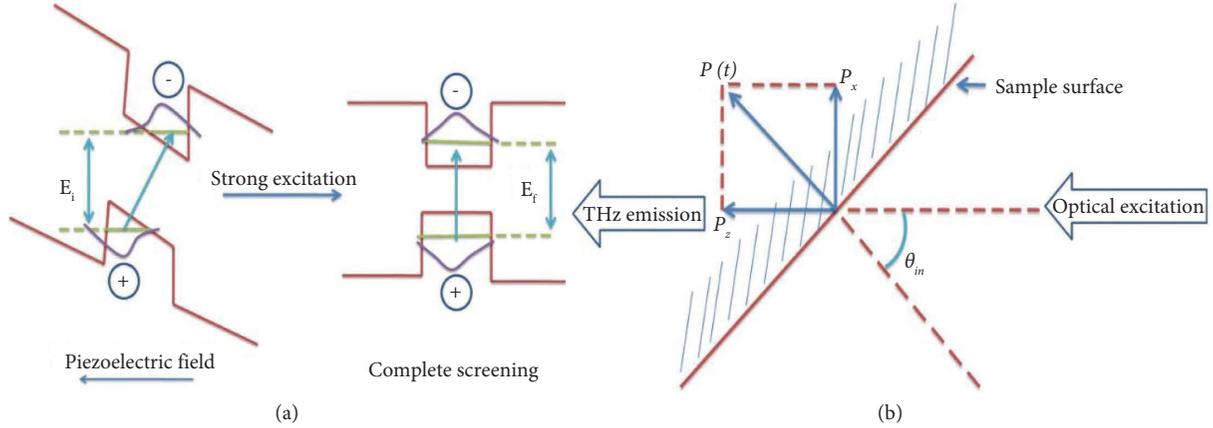

Figure 1: (a) Principle of the optically induced screening of a QW. (b) Excitation geometry for the successful THz emission from QWs.

THz emission in QWs depends on many factors such as piezoelectric field strength, QW width, period, band gaps of QWs and barriers, effective masses of electrons and holes, and band offsets. The most important parameter is the piezoelectric field strength which defines the optical and electronic properties of the QWs [25]. Here, we present a review of the most important factors that affect the THz emission from QWs under ultrafast excitation.

## 2. THz Emission Mechanism in InGaN QWs

The band tilt in QWs due to a strong piezoelectric effect is an obvious phenomenon in the absence of a restoring force. The band shape can be restored by strong excitation, which is known as screening [27, 39]. The screening effect is responsible for the THz emission in QWs under ultrafast excitation. Figure 1(a) depicts the band tilt and screening effects. The shorter time scale of the ultrafast excitation does not allow the excitons to recombine for photoluminescence (PL) emission. It rather leaves the system in a perturbed state for the subsequent photon. The absorption of the subsequent photon results in THz emission. No THz emission is observed when the sample is excited at normal incidence [33]. Excitation at $\theta_{in} = 45°$ and $-45°$ angles of incidence results in THz emission of the opposite polarities indicating its strong dependence on the piezoelectric field. Figure 1(b) is a schematic representation of the phenomenon. At this particular angle, the nonzero projection $P_x$ of the transient dipole vector can be obtained in the direction orthogonal to the propagation direction of the THz pulse [58]. This observation is clear evidence of THz emission dependence on the built-in piezoelectric field that is directed perpendicular to the sample surface. Sun et al. attained the highest THz power at an angle of incidence of 72° [59]. They suggest that the angular distribution of THz radiation is consistent with the concept of THz generation in QWs as a "radiation of dipoles."

Turchinovich et al. performed calculations that allow the derivation of polarization density $P_t(t)$ on a given instant of time in QWs under coherent photon flux $\phi(t)$ as follows [25]:

$$P_t(t) = P_o - \int_{-\infty}^{t} \frac{dP}{dt'}dt' = P_o - eL_z \int_{-\infty}^{t} \phi(t')\alpha(t')d(t')dt', \tag{1}$$

where $P_o$ is the initial polarization, e$L_z$ is the 2D array of elementary dipoles, and $\alpha$ is the absorption coefficient. $\alpha(t)$ can be derived from the following equation:

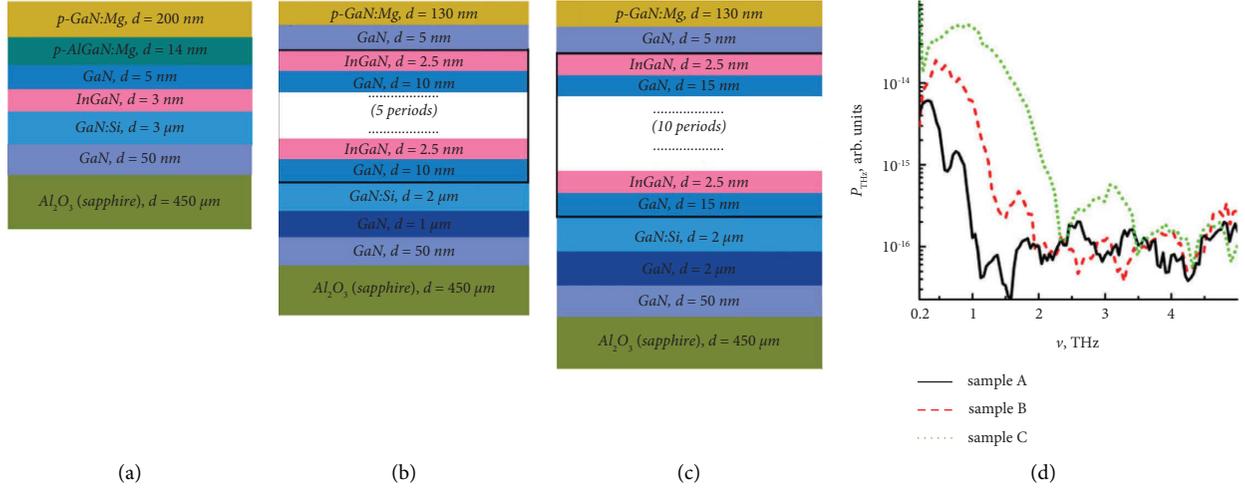

Figure 2: (a) Experimental samples with (a) one, (b) five, and (c) ten InGaN/GaN quantum wells. (d) Terahertz emission spectra generated in samples A, B, and C under the excitation power densities which provide maximal energy of the THz signal. Reprinted with permission from [26]. Reproduced with permission from Wiley materials. Copyright Wiley materials.

$$\alpha(t) = \alpha_{\max} M^2(t), \tag{2}$$

where $M(t) = \psi_e(t)|\psi_h(t)$ is the time-varying overlap integral of the empty wave function of conduction and valence band states and $\alpha_{\max}$ is the absorption coefficient of unbiased QW. The effective field $F(t)$ can be derived as a function of time by using the following equation:

$$F(t) = \frac{1}{\varepsilon \varepsilon_0 L_z} P_t(t). \tag{3}$$

## 3. Factors that Affect the THz Emission from InGaN QWs under Ultrafast Excitation

The THz emission in InGaN QWs is mainly governed by the strong built-in piezoelectric field under ultrafast excitation. QW width, period number, and Indium concentration are the main physical parameters that can influence the strength of the piezoelectric field and the resulting THz emission. The THz emission can be further tuned by changing experimental parameters such as laser energy, pump fluence, and excitation power. Here, we will discuss the most important parameters that have been reported to influence the THz emission in QWs.

### 3.1. QW Period Number.
The THz emission from InGaN QWs is strongly influenced by the number of QWs. The THz emission has been studied for QW samples with up to 16 QW period numbers [31]. It is observed that the increased period number results in an increased piezoelectric field due to strain accumulation in the respective QWs, leading to higher THz emission.

Prudaev et al. compared THz emission spectra obtained by ultrafast excitation of samples with different QW period numbers [26]. Figures 2(a)–2(c) show the structures of the samples studied. Figure 2(d) shows the emission spectra under excitation power densities that provided maximal energy for THz signals from each sample structure. With a further increase in the power densities, the THz signal was reduced. They found that an increased period number required increased excitation power densities to achieve the same maximal amplitudes of the THz pulse. The spectral maximum was found to shift to higher frequencies for samples with higher QW period numbers. THz absorption was also observed to increase with the increase in QW period number.

Sun et al. observed a continuous increase in THz output power with the increase in the QW period number [31]. They used a maximum of 16 periods. PL-saturated for period numbers above 4. The phenomenon was explained in terms of the increased piezoelectric field with the increase in QW period number due to the accumulated strain. An increased piezoelectric field reduces the wave function overlap. Therefore, the carrier recombination efficiency decreases, resulting in poor PL emission efficiencies in QWs.

### 3.2. Indium Content.
Guan Sun et al. suggested that an increase Indium content should result in higher THz emission due to increased electron-hole separation leading to higher dipole strength [31]. Norkus et al., on the other hand, found that an increase Indium content can result in the screening of built-in fields due to the high density of electron gas in QWs [41]. They explained that the ultrafast excitation of such a structure can result in the formation of a higher number of dipoles of polarity opposite to the piezoelectric field. This may suppress the piezoelectric field under increased excitation energies. It results in the saturation of THz emission efficiency for increased photon energies as shown in Figure 3(a). The screening and saturation may occur at relatively low excitation energies for such structures, limiting the extent of the THz emission.

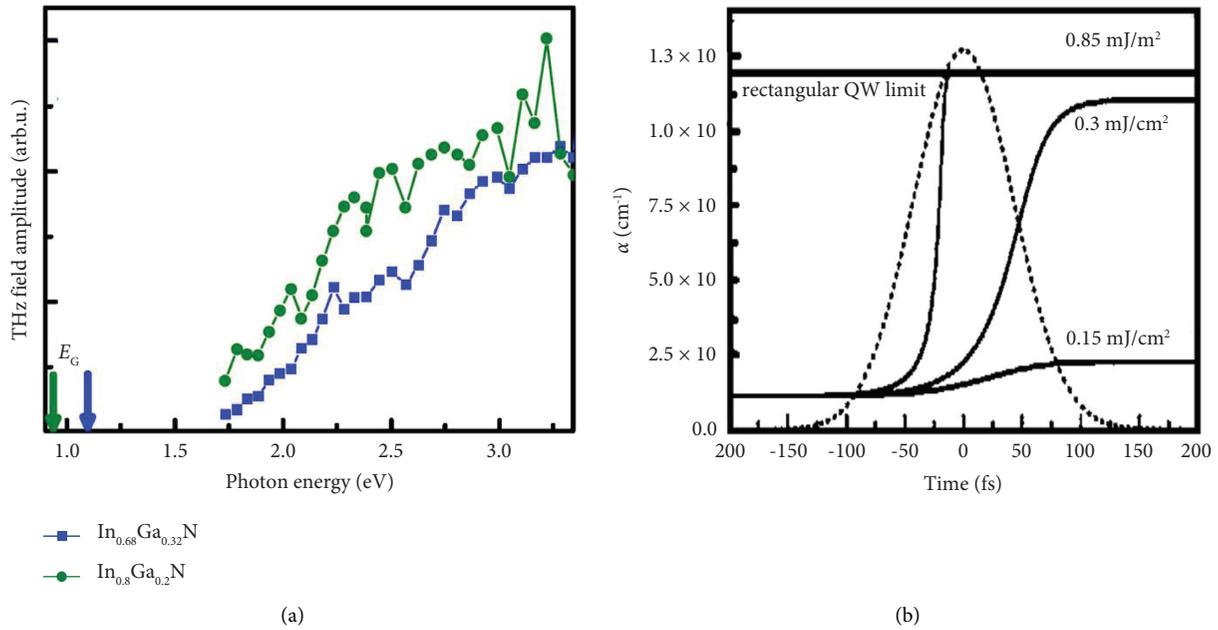

FIGURE 3: (a) THz field amplitude as a function of excitation energy in In$_x$Ga$_{1-x}$N layers with $x = 0.68$ (blue) and 0.8 (green) where $x$ represents the doping concentration. Copyright nature portfolio (b) Temporal evolution of absorption coefficient. The dashed line represents the normalized temporal shape of the photon flux in the excitation pulse. Reproduced with permission from AIP Publishing. Copyright AIP Publishing.

### 3.3. Power Density and Pump Fluence.

Excitation power densities strongly affect the THz emission in QWs. Higher emissions are observed up to certain power densities, above which the THz emission can either be reduced or saturated. The saturation is generally explained in terms of complete screening of the piezoelectric field by sufficiently high power densities. In their pioneering work, Turchinovich et al. reported THz emission from InGaN/GaN multiple QWs [58]. They found that the built-in piezoelectric field can be completely screened by the carriers excited in spatially separated states when the excitation is strong enough.

Prudaev et al. observed an increase in THz signal amplitude with an increase in power densities to a certain level [26]. For sufficiently high power densities, the THz signal amplitude was found to decrease. The increased free-carrier concentration, the increased THz absorption within the material, or the optical damage under high excitation power densities were attributed to this behavior. In previous works, saturation was observed instead [27, 30, 31, 59]. Compared with previous works, the excitation was performed at an 800 nm wavelength, when the photon energy is not enough for interband transitions to occur. The In content was also comparatively lower compared with previous works, which leads to a lower piezoelectric field.

Sun et al. observed that the increased pump fluence lead to a quadratic increase in the THz output power up to the pump fluencies of 40 $\mu$J/cm$^2$ [31]. Further increases in pump fluencies resulted in a slight deviation from a quadratic fit due to the screening effect.

The polarization dynamics in QWs for a weak and an extremely strong excitation have been studied by Turchinovich et al. [25, 58]. They calculated the excitation kinetics in a QW for variable excitation pulse fluencies, QW widths, and piezoelectric field strengths [25]. They also calculated the time evolution of the absorption coefficient for different pump fluences as shown in Figure 3(b). They found that the optical absorption coefficient is strongly affected by the excitation pulse fluence. Through calculations, they predicted the spectral broadening and shift of the THz spectra with increased excitation fluence. The lowest bandwidth limit they calculated was equal to the excitation pulse bandwidth. They suggested that such a broadening could not be detected with the conventional TDS approach, since it uses laser pulses of the same duration for both temporal shape generation and sampling. It needs to be further verified by a temporally stretched excitation and a short detection pulse.

In their experimental work, Turchinovich et al. reported the THz pulses generated in a sample with 10 identical QWs of 2.7 nm widths each [58]. The excitation fluences ranged from 0.02 to 1.3 mJ/cm$^2$. The shapes of the THz pulses were identical. They suggested that the limited bandwidth of their setup resulted in identical pulse shapes as shown in Figure 4(a). Van Capel et al. had similar observations of the THz pulse shapes as shown in Figure 4(b) [30]. For confirmation, Guan Sun et al. measured the THz spectra under different pump fluences by using a homemade submillimeter diffraction grating system [31]. The system could detect the spectral bandwidths beyond the laser spectra. They did not observe the broadening and frequency shift of the THz spectra since the screening effect was negligible in their pump fluence range. Higher fluences may allow the observation of such behavior.

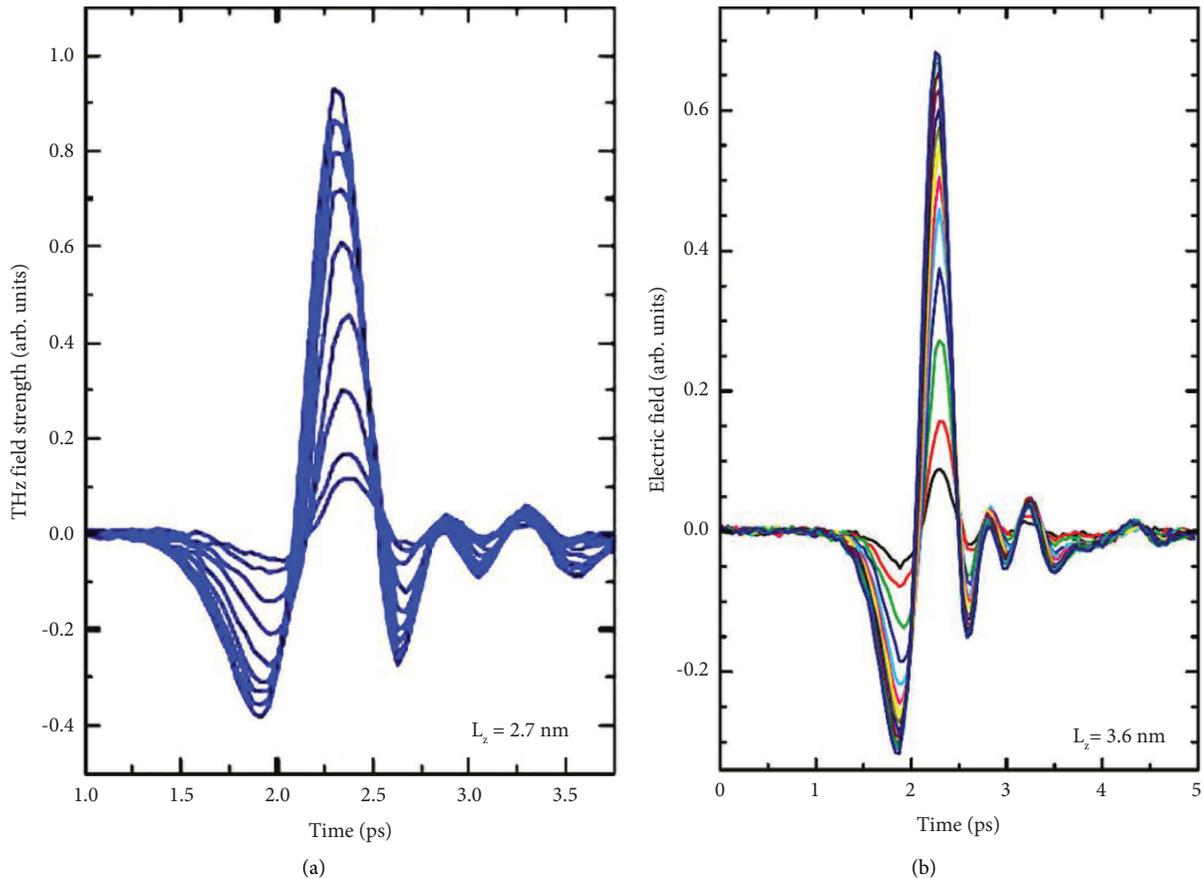

FIGURE 4: (a) THz pulses detected from samples with 2.7 nm and (b) 3.6 nm QW widths for varying pump fluences. Reproduced with permission from the American Physical Society. Copyright American Physical Society.

Above a certain pump fluence value, a saturation of THz output was observed in some works [27, 30, 31, 59]. The phenomenon is explained in terms of the depletion of the electrostatic energy in QWs by the excitation of a substantial density of polarized electron-hole pairs. The total internal reflection on the material surface and increased absorption within the material are also reported to contribute to the saturation behavior [31]. Turchinovich et al. reported the dependency of peak-peak amplitudes on the excitation fluence for two samples of 10 QWs each [58]. The QW widths were 2.7 and 3.6 nm. The pulse amplitudes increased rapidly and saturated for fluences higher than 0.5 mJ/cm², as shown in Figure 5(a). The saturation was explained by the example of a capacitor. If we consider the QW structure as a nanoscale capacitor, the maximum THz pulse energy stored in the capacitor is limited by the partial or complete discharge of the electrostatic energy stored in the nanocapacitor.

Van Capel et al. observed a monotonous increase in THz output with an increase in pump fluence and saturation above a certain fluence value, as shown in Figure 5(b) [30]. At this fluence value, complete screening of the QWs occurs, and all the energy stored in the QWs is released. For relatively lower fluence values, the screening in QWs is a negligible phenomenon [31].

### 3.4. QW Width.
Van Capel and Turchinovich et al. found that the wider QWs produced stronger pulses, as shown in Figures 5(a) and 5(b) [30, 58]. Van Capel et al. observed a 13% increase in the THz output power by increasing the QW width from 1.8 to 3.6 nm [30]. However, screening is also a dominant phenomenon in thicker QWs. Turchinovich et al., calculated the excitation kinetics in a QW for variable excitation pulse fluencies, QW widths, and piezoelectric field strengths [25]. Their calculations suggest that the thicker QWs can provide a higher number of e-h pairs with higher spatial separations, leading to higher screening in thicker QWs.

Turchinovich and Van Capel et al. found that regardless of the thickness of the QWs, saturation occurred for the same fluence value as shown in Figures 5(a) and 5(b) [30, 58]. It was explained in terms of the complete screening of the piezoelectric field that results in the release of electrostatic energy in QWs. The symmetry in QWs is restored at this moment.

### 3.5. Incident Angle.
Guan Sun et al. measured the THz output power propagating in the transmission direction for various pump beam angles (the angle of the surface normal being formed with the pump beam). They used a p-polarized pump beam for this measurement [59]. The output power was zero when the pump beam angle approached zero, whereas the THz output power was maximum for an incident beam angle of 72°. The data was well-fitted by the equation $P_{THz} \approx f(\theta)\sin^2(\theta)$, where $P_{THz}$ is the THz output power, $f(\theta)$ is the contribution from the Fresnel reflection of the pump beam, and $\sin^2(\theta)$ represents a typical angular distribution of the dipole radiation. Through this work, they proved that the angular distribution of the THz radiation is consistent with the concept of THz generation in InGaN QWs as the "radiation of dipoles" generated in the spatially separated electron and holes under the influence of a strong built-in piezoelectric field.

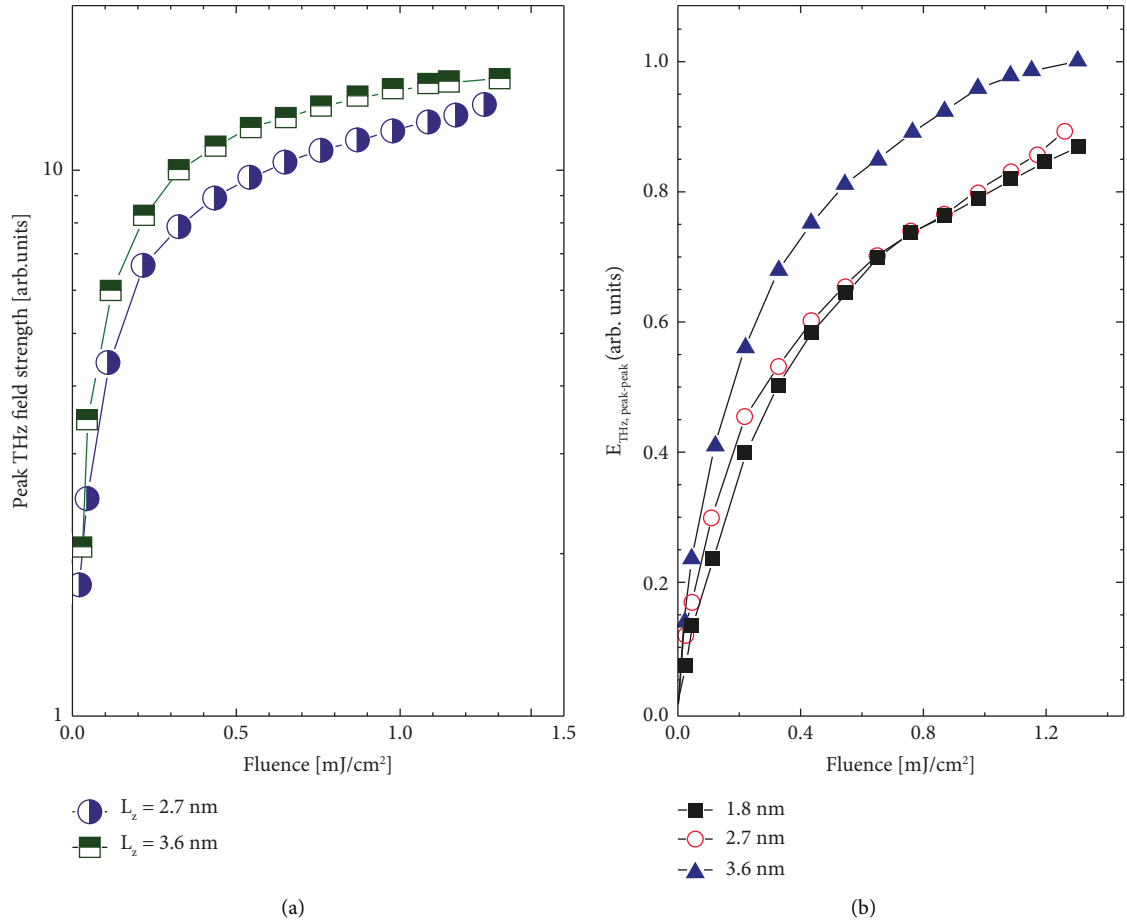

FIGURE 5: (a) Peak-peak field strengths of THz pulses for varying excitation fluences. (b) THz peak-peak amplitudes versus fluence for various QW thicknesses. Reproduced with permission from the American Physical Society. Copyright American Physical Society.

*3.6. Pump Polarization Angle.* Guan Sun et al. measured the dependences of the THz output power and polarization (angle between the THz polarization and the incident plane) on the polarization angle of the pump beam (angle between the pump polarization and the incident plane) and azimuth angle [59]. The incident angle was set to be around 72° to collect the maximum THz output power. The THz output power oscillated with the pump polarization angle. The data were well-fitted by their theoretical curve when considering the Fresnel reflection. This is primarily because of the dependence of the Fresnel reflection on the polarization angle of the pump beam. It further confirmed the concept of "radiation of dipoles" in QWs discussed above.

## 4. Correlation between THz Emission, PL, and Piezoelectric Field

The PL emission usually increases with a certain number of QWs in QW-LEDs. Further increase in period number results in saturation. Guan Sun et al. proposed that such behavior is due to the increased nonradiative structural defects with the increase in QW period number [31]. They found that THz emission is not reduced by defects because THz emission occurs as a result of absorption. On the other hand, PL emission is a radiative recombination process that is strongly affected by structural defects. It is also affected by the absorption process. The piezoelectric field also increases in QWs with the increase in period number. PL emission is strongly influenced by the piezoelectric field due to the QCSE. The QCSE is caused by band structure tilt due to the higher piezoelectric field that reduces wave function overlap. It decreases the recombination efficiencies of excitons in QWs, leading to poor PL emission efficiencies. THz emission, on the other hand, benefits from the increased piezoelectric field and lower wave function overlap. The dipoles created by spatially separated electron-hole pairs result in THz emission under ultrafast excitation.

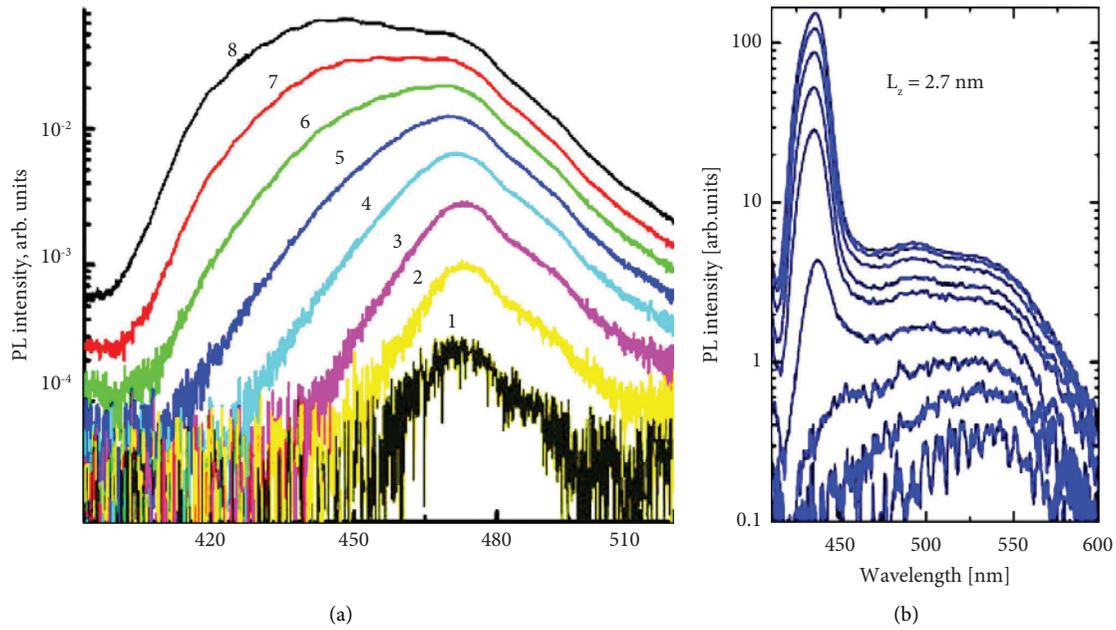

(a)

(b)

Figure 6: (a) Room temperature luminescence spectra at different excitation intensities ranging from 5.56 to 1000 kW/cm² (spectrum 1 to 8). Reproduced with permission from Wiley materials. Copyright Wiley materials (b) Time-integrated PL spectra detected from a sample with 2.7 nm QW width at excitation fluences 0.02, 0.04, 0.11, 0.22, 0.43, 0.65, 0.87, 1.09, and 1.26 mJ/cm². Reproduced with permission from the American Physical Society. Copyright American Physical Society.

Sun et al. observed that the THz output power quadratically increases with the pump fluencies up to 40 µJ/cm² [31]. A slight deviation from quadratic fit was observed for further increase in fluencies up to 85 µJ/cm². This deviation is because of the screening effect supported by the blue shift of PL observed in such experiments. However, since theoretical observations predict an increase in the absorption coefficient, a more than linear increase in PL intensity should be observed with pump fluence (the absorption coefficient is proportional to the ratio of integrated PL intensity with the pump fluence). But the PL intensities scaled up less than linearly with the pump fluencies indicating a reduction in absorption coefficients, which is not consistent with the screening effect. They concluded that the saturation behavior was because of the lower absorption coefficient at high excitation fluencies.

Ilya Prudaev et al. indicated that the local inhomogeneities in In concentration could result in an additional PL peak [26]. This additional peak, caused by asymmetric broadening and spectral shift at high excitation intensities, is attributed to the low In regions in QWs as shown in Figure 6(a). They suggested that the emission data for quantification of the piezoelectric field could thus give erroneous results. They proposed that the screening effect could not be completely understood by monitoring the PL. The localized states play an important role in the emission process that induces a band-filling effect. The band-filling effect can result in a blue shift at higher pump powers. The combination of screening and band-tail filling effects can be responsible for the observed blue shift.

Turchinovich et al. observed a broad PL peak at lower excitation fluences [58]. An additional PL peak was observed at high excitation fluences in their time-integrated PL measurements, as shown in Figure 6(b). The intensity of the additional peak increased with excitation fluences. The peak moved to 435 nm at low excitations and remained there for further increase in excitation fluences. They explained that the QW was completely biased by the built-in electric field at low excitation fluences. At high excitation intensities, the PL spectrum showed a superposition of screened (zero built-in electric fields) and biased (when all the carriers have recombined) states.

Mu et al. reported the formation of InGaN quantum dots within the InGaN QW layers [29]. They observed that the regions of high THz emission gave poor PL emissions attributed to lower densities of QDs at those locations as shown in Figure 7(a). The QDs provide photogenerated carriers localized inside them. The internal polarization field in QDs is expected to be lower compared with that in QWs. The screening of the polarization field must be higher in the areas of high QD densities, supporting the higher THz output powers. Similarly, Guan Sun et al. compared the THz and PL emissions in staggered and conventional QWs [60]. THz emission in staggered QWs was lower while the PL was higher as shown in Figures 7(b) and 7(c). Staggered QWs are designed to reduce the charge separation in QWs and increase the recombination efficiencies [61]. THz emission is obviously lower in such structures due to the lower built-in piezoelectric fields.

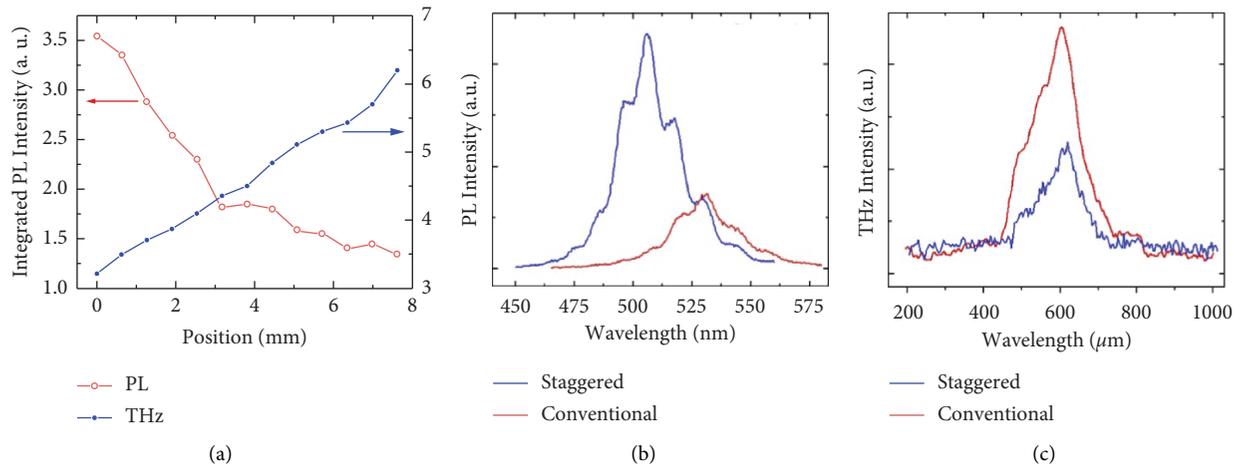

## 5. Conclusions and Outlook

The effects of certain physical and experimental parameters on THz emission in QWs have been reviewed and summarized in this work. In conclusion, an increased period number of QWs results in increased strain accumulation in the respective layers, a high piezoelectric field, increased THz emission, and increased THz absorption. Due to increased THz absorption in QW samples, emission loss within the sample may also increase. The THz frequency also shifts to higher frequencies for samples with increased QW period numbers. Increased Indium content can result in a higher number of dipoles of polarity opposite to that of the piezoelectric field under ultrafast excitation. The saturation of THz emission is an obvious phenomenon at sufficiently high excitation energies due to the complete screening of the piezoelectric field. The increased number of dipoles of opposite polarity can suppress the piezoelectric field, resulting in screening and saturation for even relatively lower excitation energies. The total internal reflection on the material surface and increased absorption within the material may also contribute to saturation. On the other hand, THz emission may also quench (instead of reaching saturation) due to the increased THz absorption within the material, optical damage under high excitation power densities, or increased free carrier concentration. Higher THz emissions can be obtained by growing thicker QWs. Although thicker QWs result in higher THz emission, saturation occurs for almost the same fluence values regardless of the thickness. The THz signal can be further optimized by calibrating the angle of incidence and pump polarization.

Mostly, the QW samples that give poor PL emission efficiencies offer higher THz emissions due to QCSE. Compared with PL, THz emission is not affected by structural defects. It may allow the use of low-quality samples and relatively less sophisticated growth techniques for low-cost THz generation from QWs. The luminescence behavior in QWs can thus predict the expected THz emission properties of QWs. Furthermore, the studies of temperature and pressure dependence on THz emission may further elaborate the emission process. The study of dipoles and magnetic moments of dipoles should lead us to an even better understanding of the phenomenon. A bigger challenge in this field may be the use of the piezoelectric field component in the growth direction to get higher TH emissions. THz emission and piezoelectric field enhancement using novel QW structures may open more opportunities in this field.

### Data Availability

No data were available.

### Conflicts of Interest

The authors declare that they have no conflicts of interest.

### Acknowledgments

This work was supported by the National Natural Science Foundation of China (61988102), Key-Area Research and Development Programof Guangdong Province (2019B090917007).